\documentclass{article}
\usepackage{amssymb}
\usepackage{bm}
\usepackage[dvips]{graphicx}
\begin{document}

\title{Spin-polarized current \\ in ferromagnetic rod-to-film structure}

\author{S. G. Chigarev, E. M. Epshtein\thanks{E-mail: epshtein36@mail.ru},
Yu. V. Gulyaev, P. E. Zilberman\\ \\
V.A. Kotelnikov Institute of Radio Engineering and Electronics\\
of the Russian Academy of Sciences, 141190 Fryazino, Russia}

\date{}

\maketitle

\abstract{Properties are discussed of a ferromagnetic junction of the type
``rod contacting with film''. Very high current density of the order of
$10^9$ A/cm$^2$ may be achieved in the contact region. We show it can lead to
inversion of population of the spin energy subbands. Spin injection
depends strongly on the direction of the current (forward or backward). We
prepared experimentally a rod-to-film structure and investigated high
density current flowing through it. Current dependent radiation has been
observed by means of a THz receiver. In particular, the radiation becomes
different for forward and backward currents. It shows the radiation
includes not only heating but also non-thermal (spin-injection) effects.}

\section{Introduction}\label{section1}
Current-induced spin injection is one of the fundamental spintronic
effects. As a very interesting application of the effect, a possibility
has been discussed of generating THz radiation by creating inverse
population of the spin subbands in a ferromagnetic layer under intense
spin injection~\cite{Kadigrobov,Gulyaev1}. By estimates, current density of the order of
$j\sim10^9$ A/cm$^2$ is needed to realize the inverse population regime in one of
the layers of a magnetic junction~\cite{Gulyaev1}.

A scheme was proposed for reaching high current density in a rod--
thin film system~\cite{Chigarev}. If the film thickness $h$ is small compared to the rod
radius $R$, then the current density in the film near the rod is $R/2h$ times
the current density in the film that allows to reach high current density
and spin injection intensity in the film. In this work, we consider
briefly the injection properties of such a structure and present
preliminary experimental results.

\section{Properties of the rod-to-film structure}\label{section2}
We consider a layered structure including two circular in plane metallic
ferromagnetic layers: layer 1 having small radius $R\le10$ $\mu$m and large enough
thickness and layer 2 having large radius and very small thickness $h\le10$ nm,
being $h\ll R$. Therefore we consider layers with very different geometric
parameters, for example, a rod contacting with a thin film.

\begin{figure}
\includegraphics{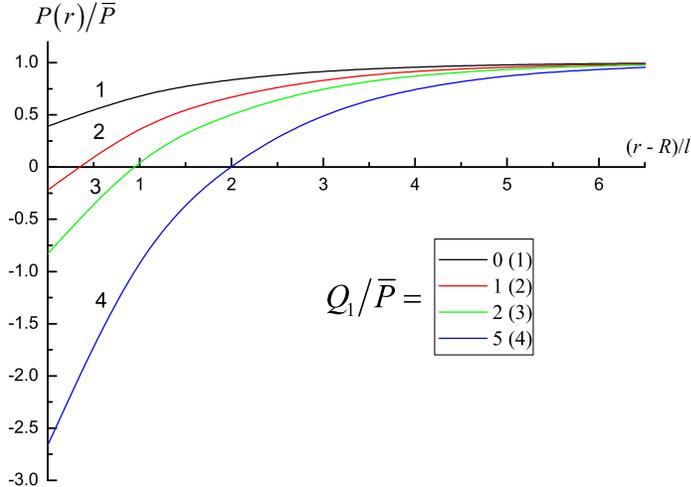}
\caption{Spatial spin polarization distribution near the rod at
$R/l=20,\,j(R)/j_D=1$ and various values of $Q_1/\bar P$ ratio.}\label{fig1}
\end{figure}

We calculate the electron spin distribution in the structure when
electrons flow in $1\to2$ direction. Detail of the calculation is in Refs.~\cite{Chigarev,Gulyaev2}.
Spin polarization $P=(n_+-n_-)/n$ in layer 2 satisfies the continuity equation
\begin{equation}\label{1}
  \nabla^2P-\frac{(\mathbf j\nabla)P}{j_Dl}-\frac{P-\bar P}{l^2}=0,
\end{equation}
where $n_\pm$ are the electron densities in low and high energy spin subbands,
respectively, $n$ is the total electron density, which is constant in the
layer due to the charge neutrality condition, $\bar P$ is the equilibrium spin
polarization, $\mathbf j$ is the current density, $j_D=enl/\tau$ is a characteristic diffusion
current density, $e$ is the electron charge, $l=\sqrt{D\tau}$ is the spin diffusion length, $D$ is
the spin diffusion constant, $\tau$ is the spin relaxation time.
Equation~(\ref{1}) was solved analytically using cylindrical coordinates with the conditions
of charge and spin currents continuity at the boundaries of the layers.
Very high current density of the order of $10^9$ A/cm$^2$ appears near the boundary
between layers 1 and 2, which exceeds by $R/2h$ times the current in the layer
1. The inversion of spin population may be created in layer 2 due to the
spin injection by current from layer 1.

\begin{figure}
\includegraphics{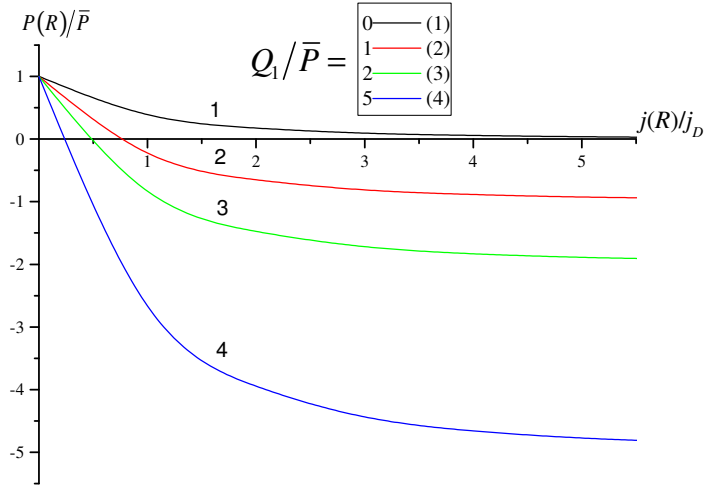}
\caption{Spin polarization at the boundary between the rod and the film
as a function of the (dimensionless) current density $j(R)/j_D$ at
$R/l=20$ and various values of $Q_1/\bar P$ ratio.}\label{fig2}
\end{figure}

The resulting spin polarization $P(r)$ tends to the equilibrium value $\bar P$ in the
film, when we are moving far from the rod edge. Calculated spatial
distribution of the polarization is shown in Fig.~\ref{fig1} at several values of
$Q_1/\bar P$ ratio, where
\begin{equation}\label{2}
  Q_1=\frac{\sigma_+-\sigma_-}{\sigma_++\sigma_-},
\end{equation}
is the spin polarization of conductivity in layer 1. We see that the
inversion of spin population ($P(R)<0$) may appear at large enough values of that
ratio.

\begin{figure}
\includegraphics{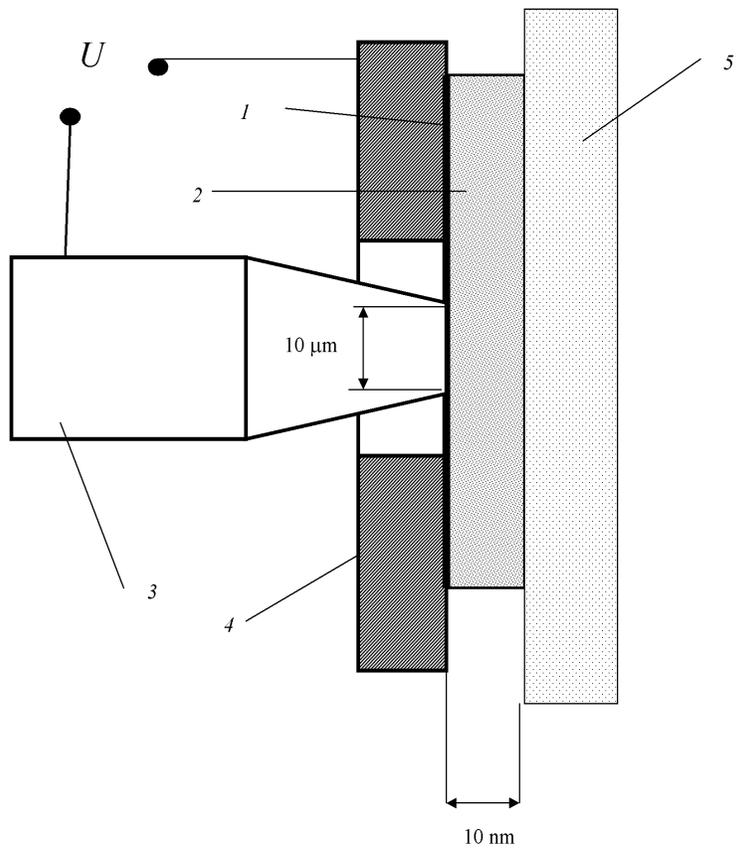}
\caption{Scheme of the rod--film emitter: \emph{1}--~ferromagnetic film, \emph{2}--~substrate,
\emph{3}--~current-carrying ferromagnetic rod, \emph{4}--~nonmagnetic current-carrying lead,
\emph{5}--~fluoroplastic plate, $U$--~voltage of the power source.}\label{fig3}
\end{figure}

The calculated dependence of the relative spin polarization $P(R)/\bar P$ on the
relative current density $j/j_D$ near the boundary between layers 1 and 2 is
shown in Fig.~\ref{fig2}. Curves 1--4 correspond to rising spin injection by
current. We see the inversion of spin population $P(R)<0$ may be achieved. Such a
possibility was discussed previously in a number of works~\cite{Gulyaev1,Osipov,Viglin}.

The highest (in magnitude) negative value of the nonequilibrium spin
polarization achieved at the boundary of the rod was obtained in the form
(see Eq. (20) in~\cite{Gulyaev2})
\begin{equation}\label{3}
  |\Delta P|=\left|\left[Q_1\left(\hat{\mathbf M}_1\cdot\hat{\mathbf
  M}(R)\right)-\bar
  P\right]\frac{j(R)}{j_D}\frac{K_\nu(R/l)}{K_{\nu+1}(R/l)}\right|,
\end{equation}
where $\hat{\mathbf M}_1,\hat{\mathbf M}$ are the unit magnetization vectors in layers 1 and 2,
respectively, $K_\nu$ is the modified Bessel function of the second kind with
index
\begin{equation}\label{4}
  \nu=\frac{1}{2}\frac{R}{l}\frac{j(R)}{j_D}.
\end{equation}
The most significant consequence of the formulae~(\ref{3}) and~(\ref{4}) is the fact
that the nonequilibrium polarization $\Delta P$ depends on the current both directly and via the
index $\nu$, being nonsymmetrical with respect to changing the current
sign, $j\to-j$. Therefore the spin-injection contributes to $\Delta P$, and the contribution is
different for forward and backward directions of the current.

\begin{figure}
\includegraphics{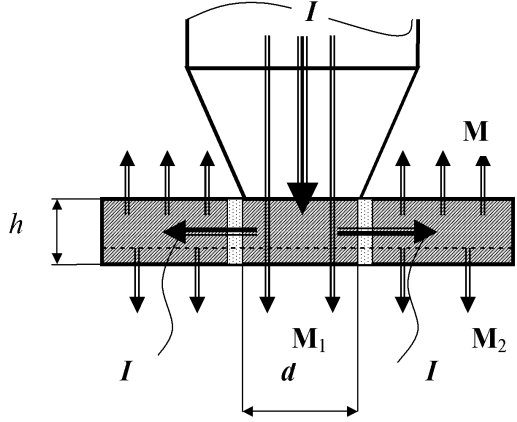}
\caption{Magnetization distribution and current direction in the rod--film system:
$I$--~current, $\mathbf{M}_1$--~the rod magnetization, $\mathbf{M},\,\mathbf{M}_2$--~magnetizations
 of the various regions of the film.}\label{fig4}
\end{figure}

\section{Experimental structure}\label{section3}
We designed a structure (Fig.~\ref{fig3}) consisted of a metallic ferromagnetic rod
\emph{3} and film \emph{1} contacted each other. This structure will be named further as
a ``rod-to-film'' one. It is simply a kind of a magnetic junction which has
specific properties. The current-carrying rod was taken of hardened steel
having approximately circular cylindrical form and minimal radius $R\approx10$ $\mu$m. In
contrast to the rod, the film part of the structure was taken magnetically
soft, for example, a film of Permalloy. Moreover, the thickness of the
film was small enough, namely, $h\approx10$ nm. The film was deposited onto a dielectric
substrate \emph{2}. The other current-carrying lead was a copper plate \emph{4} which
the magnetic film was pressed to by means a fluoroplastic plate \emph{5}. The rod
was contacting with the film through a hole in the copper plate.

As mentioned above, such a construction allows to obtain high current density (more than $10^8$ A/cm$^2$).

Based on some model experiments, a preliminary conclusion was made about
the magnetization distribution in the Permalloy film (Fig.~\ref{fig4}). The rod has
its own magnetization $\mathbf{M}_1$  along the axis that leads to spin polarization of
the current. Due to low coercivity of the film, its magnetization reverses
by the rod magnetic field. The part of the film near the surface
contacting with the rod has magnetization $\mathbf M$ with direction opposite to $\mathbf{M}_1$,
while the part near the other surface has magnetization $\mathbf{M}_2$  along $\mathbf{M}_1$.
Therefore, a situation may be obtained where the spin-polarized electrons
entering to the film from the rod find themselves in a region with
antiparallel magnetization, i.\,e., in the higher spin subband. This can
allow creating inverse spin population in the film, so that the structure
can work as a spin-injection emitter.

\begin{figure}
\includegraphics{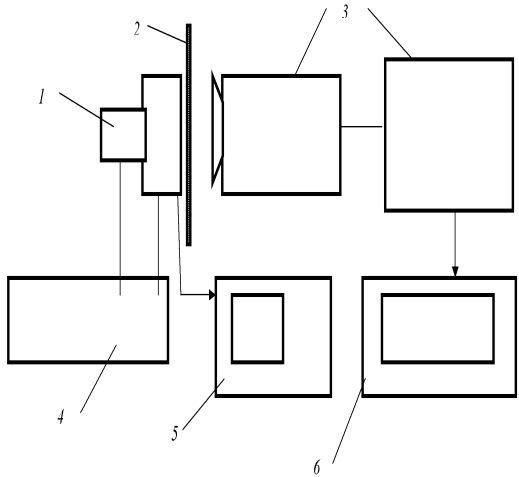}
\caption{Scheme of the experimental stand: \emph{1}--~emitter, \emph{2}--~low-frequency
filter, \emph{3}--~Golay cell with control block, \emph{4}--~pulsed voltage source, \emph{5}--~
digital oscillograph, \emph{6}--~digital recorder.}\label{fig5}
\end{figure}

\section{Measurements and results}\label{section4}
We measured radiation from the structure with a stand shown in Fig.~\ref{fig5}. The
signal from emitter \emph{1} passed from the rod--film contact through the
dielectric substrate and low-frequency filter \emph{2} was detected with Golay
cell \emph{3}. The low-frequency filter in form of a metal grid with meshes of
$125\times125$ $\mu$m$^2$ was used to cut off long-wavelength signals, because the
Golay cell can detect signals in a wide wavelength range from 10
$\mu$m to 8 mm. A pulse generator \emph{4}, used as a current source, allowed pulsed current
flowing through the system with pulse amplitude up to 0.8 A and various
pulse durations and repeating frequencies, so that the pulse
period-to-pulse duration ratio (PPPDR) was varied from 2 to 20. As a
result, current density of $2\times10^8$ A/cm$^2$ could be created in the operating range
without breaking the system. The current pulse parameters were measured by
a digital oscillograph \emph{5}, while the time depending radiation intensity was
registered by a digital recorder \emph{6}.

\begin{figure}
\includegraphics{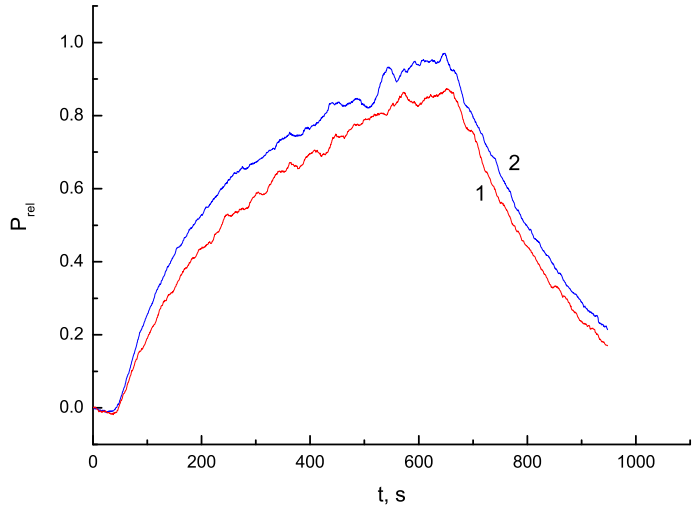}
\caption{Time dependence of the radiation intensity under forward (1) and
backward (2) current directions at PPPDR = 5.}\label{fig6}
\end{figure}

\begin{figure}
\includegraphics{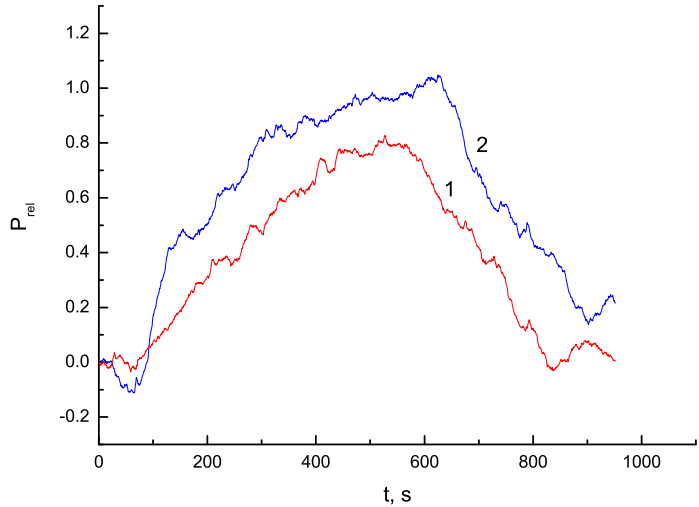}
\caption{Time dependence of the radiation intensity under forward (1) and
backward (2) current directions at PPPDR = 10.}\label{fig7}
\end{figure}

\begin{figure}
\includegraphics{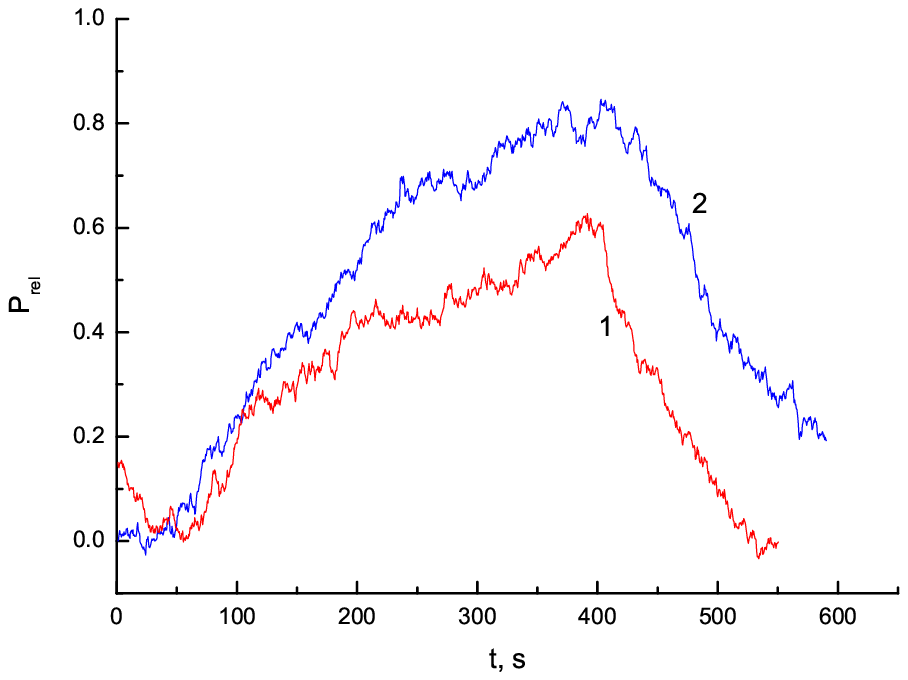}
\caption{Time dependence of the radiation intensity under forward (1) and
backward (2) current directions at PPPDR = 20.}\label{fig8}
\end{figure}

\begin{figure}
\includegraphics{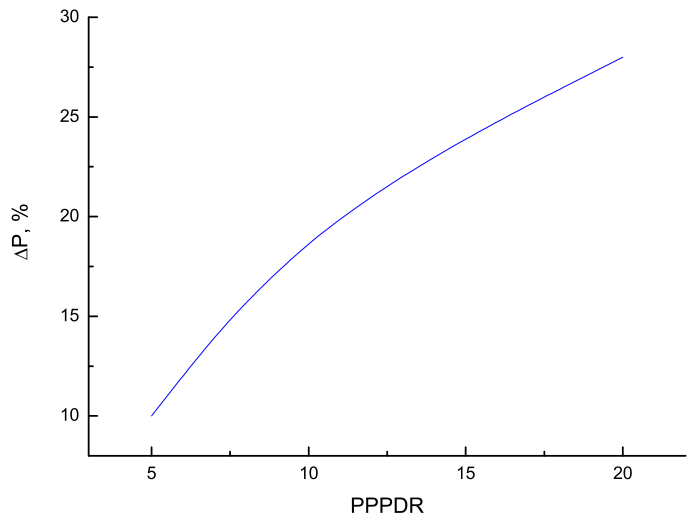}
\caption{Maximal difference in the radiation intensities between forward
and backward currents as a function of PPPDR value.}\label{fig9}
\end{figure}

Measurements were made of the radiation intensity time dependence under
forward and backward current flowing. Various PPPDR values were used. The
results are shown in Figs.~\ref{fig6}--\ref{fig8}. A summary curve in Fig.~\ref{fig9} shows maximal
difference in the radiation intensities between forward and backward
currents as a function of PPPDR value. It is seen that the measured
intensity depends on the current direction. The difference increases with
increasing PPPDR, i.\,e. with decreasing heating effect. The difference
disappeared when the magnetic (steel) rod was changed with nonmagnetic
(copper) one. This allows to
suppose that we have to deal with a non-thermal effect of the current
here. Indeed, if the observed radiation is a sum of the thermal and
non-thermal ones, then the thermal radiation intensity must decrease with
PPPDR increasing, while the non-thermal one must remain fixed. The
total intensity decreases, but the ratio of the non-thermal radiation
intensity to the thermal one must increase, as we observe.

The results obtained are unlikely may be related with thermoelectric and
thermomagnetic phenomena such as the Peltier and Ettingshausen effects.
For metals, these effects may be estimate as a fraction of a degree, while
we have heating up to 10--15 degrees.

\section{Discussion}\label{section5}
Two interesting facts have been observed in our measurements, namely,
presence of a non-thermal contribution to THz radiation from the system in
study and difference between the radiation intensities under forward and
backward current direction. It was shown earlier~\cite{Gulyaev3} that the
current-induced spin injection from a ferromagnetic layer to another one
depended substantially on the current polarity. Therefore, the facts
mentioned may be due to the radiation created by the nonequilibrium spins
near the rod--film boundary. However, more detail measurements are needed
to validate this assumption. We plan to continue this work.

\section*{Acknowledgment}
The authors are grateful to Prof. G.M. Mikhailov, Dr. V.N. Kornienko
and Dr. G.A. Ovsyannikov for valuable discussions.

The work was supported by the Russian Foundation for Basic Research,
Grants Nos.~08-07-00290 and 10-02-00039.

\end{document}